\documentclass[aps,prl,preprint,tightenlines,superscriptaddress,showpacs,
byrevtex,floatfix]{revtex4}
\usepackage{dcolumn,booktabs,threeparttable,bm}  % Align table columns on decimal point

\usepackage{epsfig}% Include figure files

\newcommand{\mb}{{M_{\rm bc}}}
\newcommand{\de}{{\Delta{E}}}

\newcommand{\pp}{{p\bar{p}}}
\newcommand{\LL}{{\Lambda\bar{\Lambda}}}
\newcommand{\LLK}{{\Lambda\bar{\Lambda}K^+}}
\newcommand{\ppk}{{p\bar{p}K^+}}

\begin{document}
\preprint{\vbox{ \hbox{   }
%                  \hbox{draft V0}
%                  \hbox{Belle Note 827}
                  \hbox{Belle Preprint 2006-18}
                  \hbox{KEK Preprint   2006-15}
%                 \hbox{EPS Parallel Sessions: 3, 10, 12}
%                 \hbox{EPS-ID 531}
%                 \hbox{hep-ex/0302024}
}}

\title{ \quad\\[0.5cm]
\boldmath Study of $J/\psi \to \pp, \LL$ and observation of $\eta_c \to 
   \LL$ at Belle}

%% >>>> Insert final authorlist here

%%% Paper:    J/psi -> p pbar, Lambda Lambdabar, eta_c -> Lambda Lambdabar
%%% Journal:  Physical Review Letters
%%% Contacts: M.Z. Wang (mwang@phys.ntu.edu.tw)
%%%           C.H. Wu (steven@hep1.phys.ntu.edu.tw)
%%% Non-responding authors or those who said NO are commented out.
%%% ====================================================================
%%% Click the RELOAD button on your web browser to see the updated file.
%%% ====================================================================
%%% Use \input{author} to insert this material into your latex file.
%%%%% Force institutions to appear in alphabetical order when typeset.
\affiliation{Budker Institute of Nuclear Physics, Novosibirsk}
\affiliation{Chiba University, Chiba}
\affiliation{Chonnam National University, Kwangju}
\affiliation{University of Cincinnati, Cincinnati, Ohio 45221}
%%%\affiliation{University of Frankfurt, Frankfurt}
%%%\affiliation{Gyeongsang National University, Chinju}
\affiliation{University of Hawaii, Honolulu, Hawaii 96822}
\affiliation{High Energy Accelerator Research Organization (KEK), Tsukuba}
%%%\affiliation{Hiroshima Institute of Technology, Hiroshima}
\affiliation{University of Illinois at Urbana-Champaign, Urbana, Illinois 61801}
%%%\affiliation{Institute of High Energy Physics, Chinese Academy of Sciences, Beijing}
\affiliation{Institute of High Energy Physics, Vienna}
\affiliation{Institute of High Energy Physics, Protvino}
\affiliation{Institute for Theoretical and Experimental Physics, Moscow}
\affiliation{J. Stefan Institute, Ljubljana}
\affiliation{Kanagawa University, Yokohama}
\affiliation{Korea University, Seoul}
%%%\affiliation{Kyoto University, Kyoto}
\affiliation{Kyungpook National University, Taegu}
\affiliation{Swiss Federal Institute of Technology of Lausanne, EPFL, Lausanne}
\affiliation{University of Ljubljana, Ljubljana}
%%%\affiliation{University of Maribor, Maribor}
\affiliation{University of Melbourne, Victoria}
\affiliation{Nagoya University, Nagoya}
\affiliation{Nara Women's University, Nara}
\affiliation{National Central University, Chung-li}
\affiliation{National United University, Miao Li}
\affiliation{Department of Physics, National Taiwan University, Taipei}
\affiliation{H. Niewodniczanski Institute of Nuclear Physics, Krakow}
\affiliation{Nippon Dental University, Niigata}
\affiliation{Niigata University, Niigata}
\affiliation{University of Nova Gorica, Nova Gorica}
\affiliation{Osaka City University, Osaka}
\affiliation{Osaka University, Osaka}
\affiliation{Panjab University, Chandigarh}
\affiliation{Peking University, Beijing}
%%%\affiliation{University of Pittsburgh, Pittsburgh, Pennsylvania 15260}
\affiliation{Princeton University, Princeton, New Jersey 08544}
\affiliation{RIKEN BNL Research Center, Upton, New York 11973}
%%%\affiliation{Saga University, Saga}
\affiliation{University of Science and Technology of China, Hefei}
\affiliation{Seoul National University, Seoul}
%%%\affiliation{Shinshu University, Nagano}
\affiliation{Sungkyunkwan University, Suwon}
\affiliation{University of Sydney, Sydney NSW}
\affiliation{Tata Institute of Fundamental Research, Bombay}
\affiliation{Toho University, Funabashi}
%%%\affiliation{Tohoku Gakuin University, Tagajo}
\affiliation{Tohoku University, Sendai}
\affiliation{Department of Physics, University of Tokyo, Tokyo}
\affiliation{Tokyo Institute of Technology, Tokyo}
\affiliation{Tokyo Metropolitan University, Tokyo}
\affiliation{Tokyo University of Agriculture and Technology, Tokyo}
%%%\affiliation{Toyama National College of Maritime Technology, Toyama}
%%%\affiliation{University of Tsukuba, Tsukuba}
\affiliation{Virginia Polytechnic Institute and State University, Blacksburg, Virginia 24061}
\affiliation{Yonsei University, Seoul}
   \author{C.-H.~Wu}\affiliation{Department of Physics, National Taiwan University, Taipei} % Taiwan
   \author{M.-Z.~Wang}\affiliation{Department of Physics, National Taiwan University, Taipei} % Taiwan
   \author{K.~Abe}\affiliation{High Energy Accelerator Research Organization (KEK), Tsukuba} % KEK
% \author{K.~Abe}\affiliation{Tohoku Gakuin University, Tagajo} % TohokuGakuin
% \author{N.~Abe}\affiliation{Tokyo Institute of Technology, Tokyo} % TIT
   \author{I.~Adachi}\affiliation{High Energy Accelerator Research Organization (KEK), Tsukuba} % KEK
   \author{H.~Aihara}\affiliation{Department of Physics, University of Tokyo, Tokyo} % Tokyo
% \author{D.~Anipko}\affiliation{Budker Institute of Nuclear Physics, Novosibirsk} % BINP
% \author{K.~Aoki}\affiliation{Nagoya University, Nagoya} % Nagoya
% \author{K.~Arinstein}\affiliation{Budker Institute of Nuclear Physics, Novosibirsk} % BINP
% \author{Y.~Asano}\affiliation{University of Tsukuba, Tsukuba} % Tsukuba
% \author{T.~Aso}\affiliation{Toyama National College of Maritime Technology, Toyama} % Toyama
   \author{V.~Aulchenko}\affiliation{Budker Institute of Nuclear Physics, Novosibirsk} % BINP
   \author{T.~Aushev}\affiliation{Institute for Theoretical and Experimental Physics, Moscow} % ITEP
% \author{T.~Aziz}\affiliation{Tata Institute of Fundamental Research, Bombay} % Tata
   \author{S.~Bahinipati}\affiliation{University of Cincinnati, Cincinnati, Ohio 45221} % Cincinnati
   \author{A.~M.~Bakich}\affiliation{University of Sydney, Sydney NSW} % Sydney
   \author{V.~Balagura}\affiliation{Institute for Theoretical and Experimental Physics, Moscow} % ITEP
% \author{Y.~Ban}\affiliation{Peking University, Beijing} % Peking
% \author{S.~Banerjee}\affiliation{Tata Institute of Fundamental Research, Bombay} % Tata
% \author{E.~Barberio}\affiliation{University of Melbourne, Victoria} % Melbourne
% \author{M.~Barbero}\affiliation{University of Hawaii, Honolulu, Hawaii 96822} % Hawaii
% ...
% USTC
% \author{V.~Zhilich}\affiliation{Budker Institute of Nuclear Physics, Novosibirsk} % BINP
% \author{T.~Ziegler}\affiliation{Princeton University, Princeton, New Jersey 08544} % Princeton
% \author{D.~Z\"urcher}\affiliation{Swiss Federal Institute of Technology of Lausanne, EPFL, Lausanne} % Lausanne
%\collaboration{The Belle Collaboration}

\collaboration{The Belle Collaboration}

%\collaboration{Belle Collaboration}
%\noaffiliation

\begin{abstract}
 
We study the baryonic charmonium decays of $B$ mesons, $B^+ \to \eta_c K^+$ 
and $B^+ \to J/\psi K^+$, where the $\eta_c$ and $J/\psi$ subsequently
decay into a $p\bar p$ or $\Lambda\bar\Lambda$ pair.  
%---------------------------------------------------------------------------
%We study the following baryonic decays of $B$ mesons: $B^+ \to \pp K^+$ and
%$B^+ \to \LL K^+$. The invariant mass spectra of the baryon-antibaryon pairs
%are dominated by $\eta_c$ and $J/\psi$ productions above the 
%charmonium threshold.   
%The $J/\psi$'s  produced in the above $B$ meson decays are in 
%a pure helicity $0$ state. 
We measure the $J/\psi \to \pp$ and $\LL$ anisotropy parameters,
$\alpha_B = -0.60 \pm 0.13 \pm 0.14 $ ($\pp$), $ -0.44 \pm  0.51 \pm
0.31$ ($\LL$) and compare to results
from $e^+e^- \to J/\psi$ formation experiments.
%----------------------------------------------------------------------------
%kichimi's suggestion
%We study baryonic B decays, $B^+ \to \pp K^+$ and $B^+ \to \LL K^+$, which are
%proceeded dominantly through two-body states of $J/\psi K^+$ and $\eta_c K^+$.
%As the helicity of the $J/\psi$ mesons is restricted to 0 in these
%decays, we determine the anisotropy parameters of $\alpha_B =
%-0.54 \pm 0.14 \pm 0.13 $ ($\pp$), $ -0.63 \pm  0.46 \pm
%0.27$ ($\LL$), and compare them with the corresponding parameters
%measured in $e^+e^- \to J/\psi$
%formation experiments.
%The polar angular distributions
%of the baryon-antibaryon pairs from $J/\psi$ decays are presented, 
%along with fit results to a
%$1 + \alpha_B\cos^2\theta$ parametrization. 
%-----------------------------------------------------------------------------
%Comparisons are made with the
%results from $e^+e^- \to J/\psi$ formation experiments. 
%-----------------------------------------------------------------------------
We also report the first
observation of $\eta_c \to \LL$. The measured branching fraction is
${\mathcal B}(\eta_c \to \LL) = ( 0.87^{+0.24}_{-0.21} (stat)^{+0.09}_{-0.14} (syst)
\pm 0.27 ({\rm PDG}))\times 10^{-3}$.
This study is based on a 357 fb$^{-1}$ data sample recorded
on the $\Upsilon({\rm 4S})$ resonance with the Belle detector at
the KEKB asymmetric-energy $e^+e^-$
collider.

\end{abstract}
\maketitle
\tighten
{\renewcommand{\thefootnote}{\fnsymbol{footnote}}
\setcounter{footnote}{0}

There have been many observations of baryonic three-body $B$ 
decays in recent years~\cite{ppk,plpi,pph,LLK,ppkBABAR}.
An interesting feature of these observations is the presence of 
peaks near threshold
in the mass spectra of the baryon-antibaryon pair.
These enhancements are not likely to be resonance states,
as the baryon angular distributions are not symmetric in their respective
helicity frames~\cite{polar}.
Other visible structures in the
mass spectra arise from charmonium decays. 
It is natural to compare the baryon angular distributions from 
charmonium decays
with those in the region of the threshold enhancement. 
There is particular interest in $J/\psi \to \pp$, where the 
proton angular distribution has been studied by
DASP~\cite{DASP}, DM2~\cite{DM2}, Mark I~\cite{MarkI},
Mark II~\cite{MarkII} and
BES~\cite{BES,Bai:1998fu,Ablikim:2005cd}.
$J/\psi$ mesons from the reaction $e^+e^-\to J/\psi$ are produced predominantly in 
helicity $\pm 1$ states. Therefore, the baryon angular distributions 
are proportional to $1+\alpha \cos^2\theta$, 
where $\theta$ is the baryon polar angle
in the  $J/\psi$ helicity frame. Many theoretical 
predictions~\cite{theory} exist for the value of $\alpha$. 
%The current world average value of $\alpha$, 
%obtained with above measuremens 
%from $J/\psi \to p\bar p$ decays, is $0.66 \pm 0.05$.

Study of two-body baryonic decays of charmonia at a $B$-factory
has several different features as compared with an $e^+e^-$ machine
running at the $J/\psi$ mass. 
$J/\psi$ mesons from the two body decay of $B$ mesons accompanied 
by spin zero particles are in a pure helicity zero state. This provides 
a useful cross check for previous measurements. 
The charmonia from $B$ decays 
do not suffer from poor acceptance near the beam pipe, and
events with $|\cos\theta|$ near 1 can be detected. Such events are
very effective for determining $\alpha$.
%do not suffer from the beam hole effect, such that events
%with $|\cos\theta|$ near 1 can be detected. 
%These events are effective to determine $\alpha$. 
%A $B$-factory is also immune from 
Requiring that the $J/\psi$ originate from a $B$ decay eliminates
$e^+e^- \to q\bar{q} \to p \bar{p}$
background, where $q$ stands for a $u$ or $d$ quark. 
%For previous studies this background 
%is intrinsically embedded and hard to separate on an event-by-event 
%basis.
For $e^+e^- \to J/\psi \to \pp$, this background cannot be separated
from the signal on an event-by-event basis.
% There have been many reported observations of baryonic three-body $B$
%decays in recent years~\cite{ppk,plpi,pph,LLK}.
%An interesting feature of these observations is the presence of a
%peak near threshold
%in the mass spectra of the baryon-antibaryon pair.
%Studies show that
%these enhancements are not likely to be resonance states,
%as the baryon angular distributions are not symmetric in their respective
%helicity frames~\cite{polar}.
%Other visible structures in the
%mass spectra arise from charmonium decays.
%It is natural to compare the baryon angular distributions from
%charmonium decays
%with those in the region of the threshold enhancement.

In the study of two-body baryonic decays of charmonia we focus
on the decay processes $B^+ \to \ppk$
and $B^+ \to \LLK$~\cite{conjugate}.
We report the first observation of $\eta_c \to \LL$.
There is little information about $\eta_c$ decays into
baryon-antibaryon pairs except for $\eta_c \to \pp$. Measuring decay rates of
the $\eta_c$ to different baryon-antibaryon modes is a
useful check for theoretical predictions~\cite{etactoBB} and
can shed light on quark-diquark dynamics.
%This can test the underlying quark-diquark
%scheme to model baryons.

We use a 357 fb$^{-1}$ data sample
consisting of $ 386 \times 10^6$ $B\bar{B}$ pairs
collected by the Belle detector %on the $\Upsilon({\rm 4S})$ resonance
at the KEKB asymmetric energy $e^+e^-$ (3.5 on 8~GeV) collider~\cite{KEKB}.
The Belle detector is a large solid angle magnetic spectrometer
that consists of
%a three layer silicon vertex detector (SVDI) for the first
%sample of $152\times 10^{6}$ $B\bar{B}$ pairs,
a four layer silicon vertex detector (SVD),
%for the later $234\times 10^{6}$ $B\bar{B}$ pairs,
a 50 layer central drift chamber (CDC), an array of aerogel threshold
\v{C}erenkov counters (ACC), a barrel-like arrangement of time of
flight scintillation counters (TOF), and an electromagnetic
calorimeter comprised of CsI (Tl) crystals located inside a
superconducting solenoid coil that provides a 1.5~T magnetic
field.  An iron flux return located outside of the coil is
instrumented to detect $K_L^0$ mesons and to identify muons. The
detector is described in detail elsewhere~\cite{Belle}.

%In this study of two-body baryonic decays of charmonia we focus 
%on the decay processes, $B^+ \to \ppk$
%and $B^+ \to \LLK$~\cite{conjugate}. 
The event selection criteria are based on information obtained
from the tracking system
(SVD+CDC) and the hadron identification system (CDC+ACC+TOF).
%All primary charged tracks
%are required to satisfy track quality criteria
%based on the track impact parameters relative to the
%interaction point (IP). %, which is determined run-by-run.
%The deviations from the IP position are required to be within
%$\pm$1 cm in the transverse ($x$--$y$) plane, and within $\pm$3 cm
%in the $z$ direction, where the $z$ axis is defined as the opposite
%direction to the positron beam. 
We follow the same procedure as in ref.~\cite{pph} to select proton and 
kaon candidates.
$\Lambda$ candidates are reconstructed via the $p\pi^-$ channel using the
method described in ref.~\cite{LAM}.

To identify the reconstructed $B$ meson candidates, we use the
beam energy
constrained mass $\mb = \sqrt{E^2_{\rm beam}-p^2_B}$ and the
energy difference $\de = E_B - E_{\rm beam}$, where $E_{\rm
beam}$ is the beam energy, and $p_B$ and $E_B$ are the momentum and
energy of the reconstructed $B$ meson in the rest frame of
the $\Upsilon({\rm 4S})$. The signal region is
defined as 5.2~GeV/$c^2 < \mb < 5.29$~GeV/$c^2$ and 
$ -0.1$ GeV $ < \de < 0.2$ GeV.
%From a GEANT based simulation, 
The signal peaks at $\mb$ = 5.279 GeV/$c^2$ and 
$\de$ = $0$.
% The $\mb$ resolution is about 3 MeV/$c^2$
%and $\de$ resolution is about 10 MeV.

The dominant background arises from continuum $e^+e^-
\to q\bar{q}$ processes.
%with much smaller contributions from ``cross-feed'', where similar 
%types of rare decay events pass each other's signal criteria.
The background from $b \to c$ and from $B$ decays into charmless
final states is negligible.
In the $\Upsilon({\rm 4S})$ rest frame,
continuum events are jet-like while
$B\bar{B}$ events are more spherical. 
The reconstructed momenta 
of final state particles are used to form several event shape variables (e.g. thrust
angle, Fox-Wolfram moments, etc.) in order to categorize each event.  
We follow the scheme described in ref.~\cite{etapk} that
combines seven event shape variables into
a Fisher discriminant to suppress
continuum background.

Probability density functions (PDFs) for the Fisher discriminant and
the cosine of the angle between the $B$ flight direction
and the beam direction in the $\Upsilon({\rm 4S})$ rest frame
are combined to form the signal (background)
likelihood ${\cal L}_{s (b)}$.
The signal PDFs are determined from GEANT based Monte Carlo (MC)
simulation and the background PDFs are obtained from 
sideband data %the continuum MC
%simulation for events with
with $\mb < 5.26$ GeV/$c^2$.
We require
the likelihood ratio ${\cal R} = {\cal L}_s/({\cal L}_s+{\cal L}_b)$ 
to be greater than 0.4 for both
$\ppk$ and $\LLK$ modes.
%The selection
%points are determined by optimization of $N_s/\sqrt{(N_s+N_b)}$, where $N_s$ 
%and $N_b$
%denote the number of signal and background, respectively. 
%Note that a nominal signal branching fraction~\cite{pph,LLK} is assumed
%to determine $N_s$.
These selection criteria suppress approximately 69\% (66\%) of 
the background while 
retaining 92\% (91\%) 
of the signal for the $\ppk$ ($\LLK$) mode.
%In this study only one $B$ candidate is allowed per event. 
If there are  multiple $B$ candidates in an event, we 
select the one with the best $\chi^2$ value from the $B$ decay vertex fit.
Multiple $B$ candidates are found in less than 2\% (5\%) of events
for the $\ppk$ ($\LLK$) mode. 

\begin{figure}[htb]
\epsfig{file=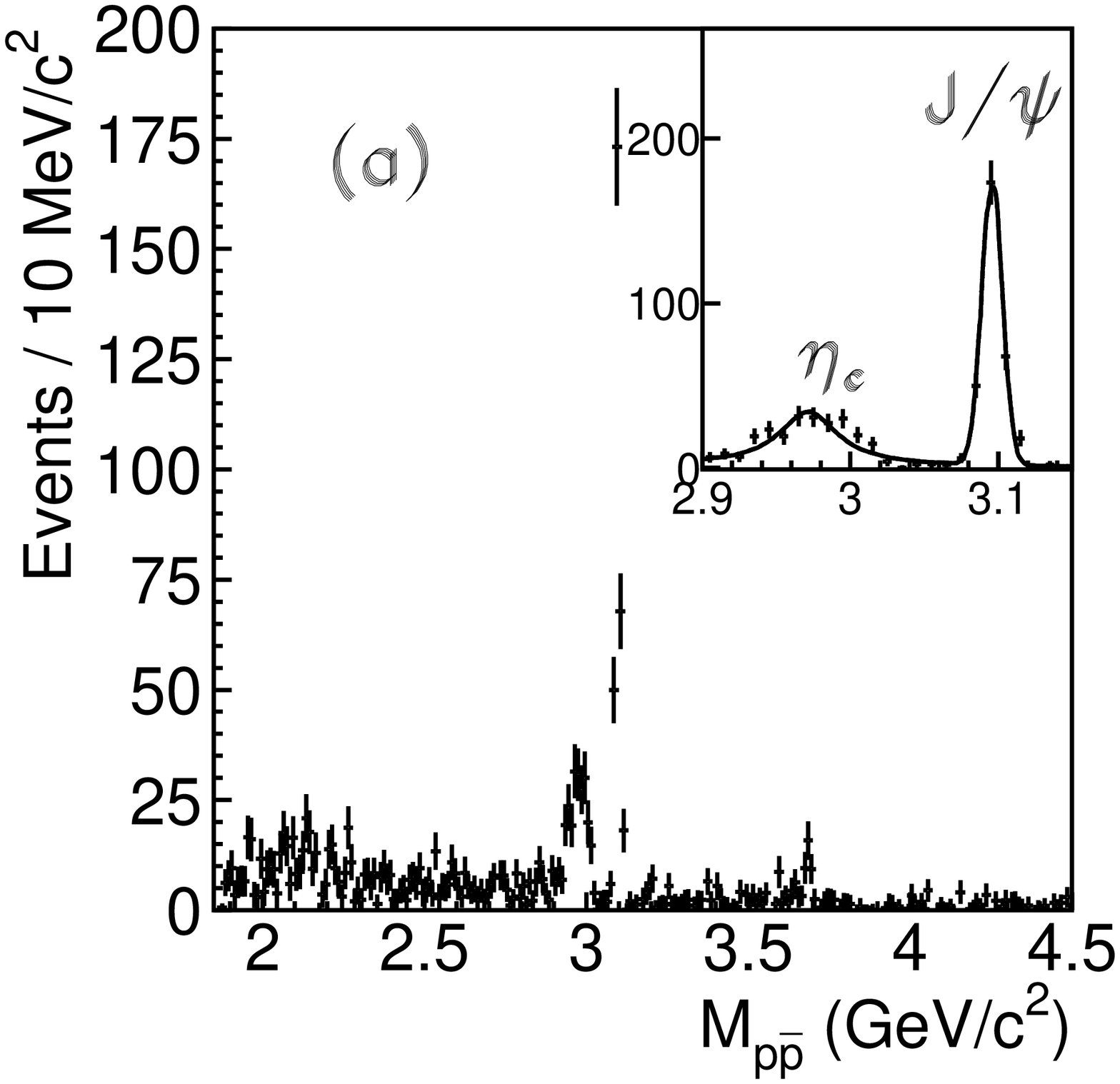,width=3in}
\epsfig{file=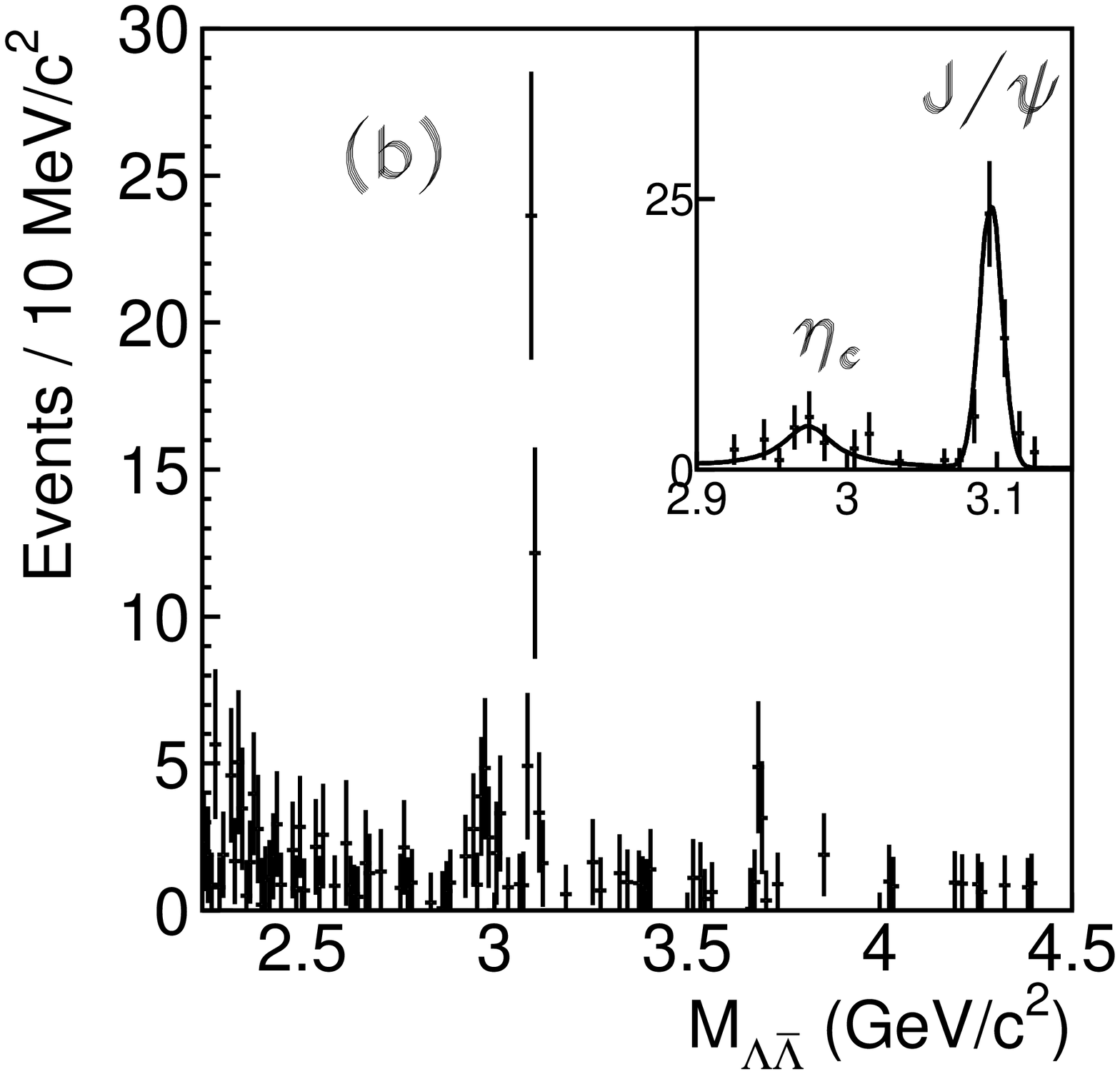,width=3in}
\caption{(a) $B$ signal yield versus  
$M_{\rm p\bar p }$ and (b) $B$ signal yield versus
$M_{\rm\Lambda\bar\Lambda}$. The inset shows the $\eta_{c}$-$J/\psi$ mass
region. The curves represent the unbinned likelihood fits to the data.}

\label{fg:mpp}
\end{figure}

We use an unbinned extended maximum likelihood fit 
to estimate the $B$ signal yield.
%:
%Events are fitted using the extended likelihood function:
%$$ L = {e^{-(N_s+N_b)} \over N!}\prod_{i=1}^{N} 
%[N_s P_s(M_{{\rm bc}_i},\Delta{E}_i)+
%N_b P_b(M_{{\rm bc}_i},\Delta{E}_i)],$$
%where $P_s(P_b)$ denotes the signal (background) PDF, 
%$N$ is the total number of events in the fit, and signal $N_s$ and background $N_b$
%yields are floated.
For the signal PDF,
we use a Gaussian in $\mb$ and a double Gaussian in $\de$.  We fix
the parameters of these functions to the values determined from 
MC simulation~\cite{correction}.
Background shapes are fixed from fitting to the sideband
events in the region 3.14 GeV/$c^2$ $ < M_\pp <$ 3.34 GeV/$c^2$.
%where $B$ signal yield is negligible.
The $\mb$ background is modeled using a parametrization used 
by the ARGUS collaboration~\cite{Argus}.
%$ f(x)\propto \mb\sqrt{1-x^2}
%\exp[-\xi (1-x^2)]$,  
%background parametrization first used by the ARGUS collaboration~\cite{Argus} 
%where $x = \mb/E_{\rm beam}$ and $\xi$ is
%fixed from sideband events. 
The $\de$ background shape is modeled by a first order polynomial.
% and a first order polynomial 
%straight line 
%for the
%$\de$ background shape. 
%There are possible cross-feeds from
%$\ppkst$ and $\ppksz$ modes to $\ppk$ ($\LLK$)
%modes; therefore the cross-feed region ($\de < -0.1$ GeV) 
%is excluded in the fit.

%significance is defined as $\sqrt{-2 {\rm ln}(L_0/L_{max})}$~\cite{PDG}, 
%where $L_0$ and
%$L_{max}$ denote the likelihood with signal yield fixed at
%zero and at the fitted value, respectively. 
%by the maximum value of the likelihood function
%and the likelihood value with yield fixed at zero~\cite{PDG}.
%The yield for $\ppksz$ is less significant.
%The non-resonant $\ks \pi^+$ component of the 
%$\ppkst$ mode is included in the
%systematic error estimation as described later.

%As the mass resolution of $M_\pp$ ($M_\LL$) is below 10 MeV/$c^2$, 
We determine $B$ signal yields %as a function of $M_\pp$ ($M_\LL$) from the 
%kinetic threshold to
%4.5 GeV/$c^2$ 
in 10 MeV/$c^2$ wide $M_\pp$ ($M_\LL$) mass bins from the
kinematic threshold to 4.5 GeV/$c^2$; the result is shown in
Fig.~\ref{fg:mpp}(a) (Fig.~\ref{fg:mpp}(b)).
There are clear $\eta_c$ and $J/\psi$ peaks in the mass spectrum.
%A fit to the data is shown in the inset.
We use a relativistic Breit-Wigner function for the $\eta_c$ peak, 
a Gaussian for the $J/\psi$ peak, and 
a linear function for the non-resonant background. The Breit-Wigner function
is convolved with the detector response function, which is taken from the
$J/\psi$ peak.
A maximum likelihood fit to the data is shown in the inset. 
%Note that the Breit-Wigner function
%is convoluted with the detector response function which is taken from the
%$J/\psi$ peak. 
%The observed mean and width of the $J/\psi$ peak is
%consistent with MC simulation. 
We obtain an $\eta_c$ mass of 
M$_{\eta_c}=2971\pm 3^{+2}_{-1}$ MeV$/c^2$ ($2974\pm 7^{+2}_{-1}$ MeV$/c^2$) and a width of 
$\Gamma (\eta_c)=48^{+8}_{-7}\pm 5 $ MeV$/c^2$ ($40 \pm 19\pm 5 $ MeV$/c^2$)
%from maximum likelihood fit
%with $246.0 \pm 28.0$ ($24.1 \pm 6.7$) signal events of 
for the $\eta_c \to \pp$ ($\eta_c \to \LL$) mode.
The systematic errors are determined from the differences of $J/\psi$ peaks 
between data and the PDG~\cite{PDG} value, and by varying different fit 
shapes for $\eta_c$ signal and background assuming
no interference effect between them.
The width is larger than the PDG average but is consistent with recent 
BaBar~\cite{Aubert:2003pt,ppkBABAR}
and previous Belle~\cite{Fang:2002gi} measurements.

\begin{figure}[htb]
\epsfig{file=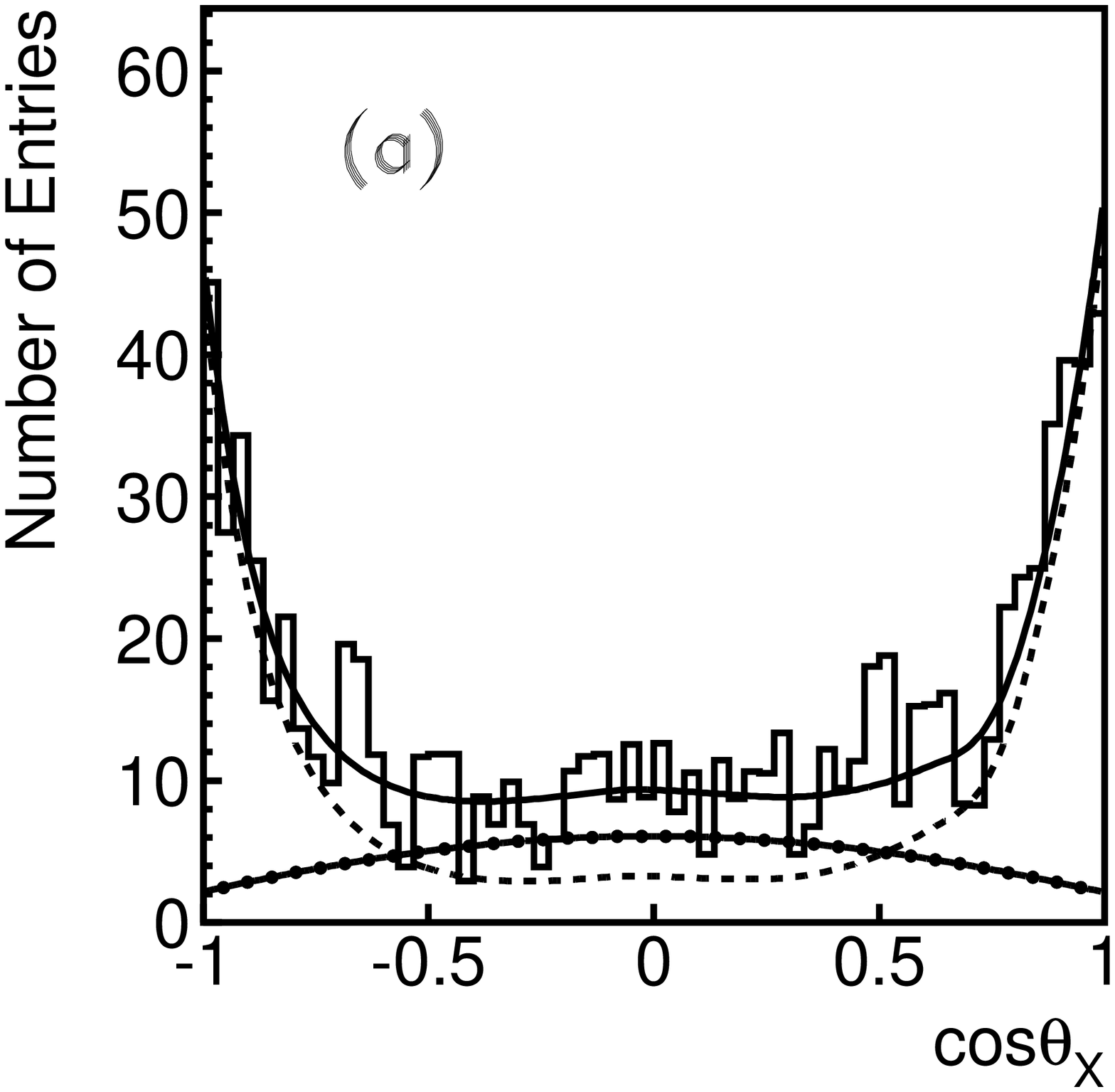,width=3in}
\epsfig{file=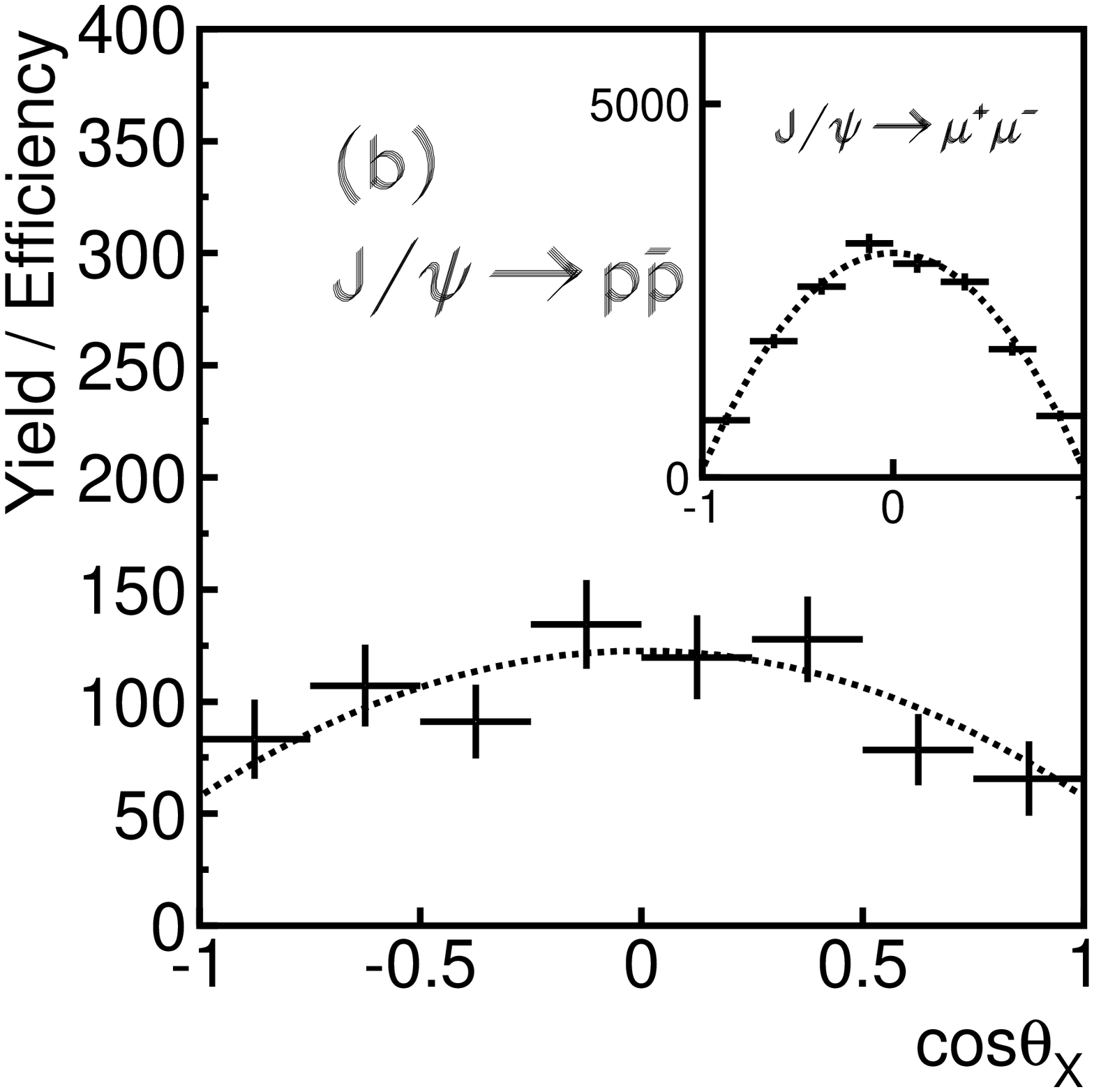,width=3in}

\caption{(a) Likelihood fit and (b) $\chi^2$ fit results of
the $J/\psi\to p\bar p$ helicity angle
distribution. In the maximum likelihood fit plot, the solid, dotted solid, 
and dashed
line represent the combined fit result, fitted signal, and fitted  background, respectively.
In the $\chi^2$ fit plot, the inset shows the fit result for $B$ signal yield
of $B^+\to J/\psi K^+, J/\psi \to \mu^+\mu^-$.}
\label{fg:jpsipp}
\end{figure}

%\begin{figure}[htb]
%\epsfig{file=fig3.eps,width=3.5in}
%\caption{The $\cos\theta_X$ distribution of $J/\psi\to p\bar p$ in the signal
%region ($\mb, \de$) and the same distribution from sideband data ($\mpp$) as
%shaded histogram.}
%\label{fg:cos-signal}
%\end{figure}

%\begin{figure}[htb]
%\epsfig{file=fig3.eps,width=3.9in}
%\caption{$J/\psi\to p\bar p$
%helicity angle distribution. The dashed line
%shows the $\chi^2$ fit result for $B$ events of
%$B^+\to J/\psi K^+, J/\psi\to p\bar p$. 
%The inset shows the $\chi^2$ fit result for
%$B$ yield of $B^+\to J/\psi K^+, J/\psi \to \mu^+\mu^-$.}
%\label{fg:comb-cos}
%\end{figure}

We define the $J/\psi$ signal region as
3.075~GeV/$c^2 < M_{\pp} < 3.117$~GeV/$c^2$ and
use events in this signal region to study the proton angular distribution 
in the helicity frame of the $J/\psi$. The helicity angle $\theta_X$ is defined as the
angle between the proton flight direction and the direction opposite
to the flight of the kaon in the $J/\psi$ rest frame.
The angular distribution of $J/\psi$ in the helicity zero state
is parameterized as $ P(\alpha_B,\cos\theta_X) = 
(1+\alpha_B \cos^2\theta_X)/(2+2\alpha_B/3)$ with $\alpha_B = {-2 \alpha/(\alpha + 1)}$~\cite{helform}.
Here, $\alpha$ is the anisotropy parameter determined
from the angular distribution of $J/\psi$ in helicity$= \pm 1$
states produced in $e^+e^- \to J/\psi$. 
Previous measurements~\cite{DASP,DM2,MarkI,MarkII,BES,Bai:1998fu,Ablikim:2005cd}
give an average of $\alpha = 0.66 \pm 0.05$ for $\pp$ and 
$\alpha = 0.65 \pm 0.11$ for $\LL$; 
these values correspond to $\alpha_B = -0.80 \pm 0.04$ for $\pp$ and $\alpha_B = -0.79 \pm 0.08$ for $\LL$. 
%as shown in TABLE~\ref{alphalist}.

%Note ethat $\alpha_B$ determined from 
%longitudinally polarized $J/\psi$ (helicity $0$) is related to 
%the $\alpha$
%determined from transversely polarized $J/\psi$ (helicity $\pm 1$)
%by $\alpha_B = {-2 \alpha \over (\alpha + 1)}$~\cite{helform}.
%From previous measurements of $\alpha$ as shown in TABLE~\ref{alphalist},
%the expectation of $\alpha_B$ is
%$-0.80 \pm 0.04$. 
%Note that after charge conjugation, 
%the angle is determined by $p$ and $K^+$ (or $\bar{p}$ and $K^-$) for
%the $\ppk$ mode. Fig.~\ref{fg:thetap} shows the efficiency corrected 
%$B$ yield in bins of $\rm{cos} \theta_X$. 

For analysis of the angular distribution, we define  a likelihood $L$,
$${e^{-(N_s+N_b)} \over N!}\prod_{i=1}^{N}
[N_s P_s(M_{{\rm bc}_i},\Delta{E}_i)
 \epsilon(\cos\theta_X{_i}) P(\alpha_B,\cos\theta_X{_i})$$
$$+
N_b P_b(M_{{\rm bc}_i},\Delta{E}_i,\cos\theta_X{_i})],$$
where $\alpha_B$ is a fit parameter in addition to $N_s$ and $N_b$,
$\epsilon(\cos\theta_X)$ is the efficiency function and $\epsilon P$
is normalized to $1$.
The efficiency $\epsilon(\cos\theta_X)$ obtained from the signal MC simulation is flat
as a function of $\cos\theta_X$.
From a study of signal MC simulation, we find that
there is no correlation between $\mb$, $\de$ and $\theta_X$. 
The background PDF as a function of $\mb$, $\de$, and $\cos\theta_X$ is 
determined from
$M_\pp$ sideband data. 
%in the region of 3.14 GeV/$c^2$ $ < M_\pp <$ 3.34 GeV/$c^2$.
Fig.~\ref{fg:jpsipp}(a) shows the result of the fit to the $J/\psi \to \pp$ 
candidates in 
the entire $\mb$, $\de$ region. We determine $\alpha_B$ to be $-0.60 \pm 0.13$.
As a cross-check, we use a $\chi^2$ method and fit 
the efficiency corrected $B$ signal yields in bins of $\cos\theta_X$ to a 
$1+\alpha_B\cos^2\theta_X$ parametrization. 
%as a function of $\cos\theta_X$. 
The results of the fit are
shown in Fig.~\ref{fg:jpsipp}(b). We obtain $\alpha_B = -0.53 \pm 0.15$ 
with $\chi^2/d.o.f. = 0.9$, consistent with the result of the unbinned fit.
%This is consistent
%with the likelihood result from a toy MC study, where samples
%with the same number of events as our data 
%are generated to check the $\alpha_B$ difference between the likelihood 
%and $\chi^2$ methods.
%The $\alpha_B$ difference follows a Gaussian
%distribution with a width of approximately $0.05$.
We measure the angular distribution of  
%We also apply 
%both the likelihood and $\chi^2$ methods to a 
$J/\psi \to \mu^+\mu^-$ decays from $B^+\to J/\psi K^+$ to verify 
the fitting procedure.
The result is shown in the inset of Fig.~\ref{fg:jpsipp}(b).
The fitted value agrees with the expectation for massless fermions.
% of $\sin^2\theta_X$.
%It is in excellent agreement
%with theoretical prediction which has a $\sin^2\theta_X$ shape ($\alpha_B
%= -0.999 \pm 0.003$).

We determine the systematic error in $\alpha_B$ by varying the value of 
various selection cuts and parameters of PDFs to check for trends in
the value of $\alpha_B$. These trends are parametrized by a linear function. 
%This relation is smooth
%and can be fitted to a linear function. 
We then quote the change in $\alpha_B$ along the line between 
the selected point
and the far end of the tested region as a systematic error. Note that
this is a conservative estimate, since statistical fluctuations
also contribute to changes in $\alpha_B$. 
We assign a systematic error of $0.08$ for the ${\cal R}$ selection,
$0.06$ for proton/kaon selection, and $0.02$ for fitting PDFs. 
Other systematic 
errors are
negligible.
%To be conservative, we also quote the observed  
%difference between the likelihood method and the $\chi^2$ method as a
%systematic error.
The observed difference between the maximum likelihood
method and the $\chi^2$ method is also included in the systematic error.
The total systematic uncertainty in $\alpha_B$ is 0.13.

%\begin{figure}[htb]
%\epsfig{file=fig4.eps,width=3.9in}
%\caption{$B$ yield versus $M_{\rm\Lambda\bar\Lambda}$. The inset shows the
%$\eta_{\rm c}$ and $J/\psi$ mass region.(blue line represents the
%non-charmonium background
%shape)}
%\label{fg:mll}
%\end{figure}

%The $\LL$ mass spectrum for 
%$B^+ \to \LLK$ decays is shown in Fig.~\ref{fg:mpp}(b).
%Similar structures are seen in the mass spectrum of $B^+ \to \LLK$ decays
%as were seen in the $B^+ \to \ppk$ mass spectrum.
There are several complicating factors in the analysis
of $B^+ \to \LLK$ decays relative to $B^+ \to \ppk$ decays.
The efficiency for detecting slow pion from $\Lambda$ decays is small.
As a result, the $\Lambda$ reconstruction efficiency is non-uniform 
as a function of polar angle ($\cos\theta_{p}$) of the secondary proton 
in the $\Lambda$ helicity frame, and is correlated with $\cos\theta_X$, 
where $X$ refers to the $\Lambda$.
The likelihood function is similar to the previous one except that the angular part contains
two more variables, $\cos\theta_{p}$ and $\cos\theta_{{\bar p}}$.    
The efficiency function 
$\epsilon(\cos\theta_X,\cos\theta_p,\cos\theta_{\bar p})$ 
is obtained from a signal MC sample with $ 4 \times 10^6$ events.
The background PDF is determined from
$M_\LL$ sideband data in the region 3.14 GeV/$c^2$ $ < M_\LL <$ 3.54 GeV/$c^2$.
The value of $\alpha_B$ obtained from the maximum likelihood fit is $-0.44 \pm 0.51 \pm 0.31$, 
where the systematic error is determined from the same procedure as that used 
for $J/\psi \to \pp$ decays.

%The $\mb$ distribution (with $|\de|<$ 0.05
%GeV) and the $\de$ distribution (with $\mb >$ 5.27 GeV/$c^2$)
%for $B^+ \to \eta_c K^+, \eta_c \to \LL$ decays are
%shown in Fig.~\ref{fg:etac_2dp}.
We define an $\eta_c$ signal region as 2.94~GeV/$c^2 < M_{\LL} < 3.02$~GeV/$c^2$. 
Signal peaks are visible in the $\mb$ and $\de$ distributions.
The fitted $B$ signal yield, efficiency, and obtained branching fraction
are shown in Table~\ref{br_charm}.
The maximum likelihood fit for $B^+\to\eta_{\rm c}K^+,\eta_{\rm c}\to\Lambda\bar\Lambda$ gives a yield of $ 19.5^{+5.1}_{-4.4}$ 
with a statistical significance of $7.9$ standard deviations.
The significance is defined as $\sqrt{-2{\rm ln}(L_0/L_{\rm max})}$,
where $L_0$ and
$L_{\rm max}$ are the likelihood values returned by the fit with
the signal yield fixed to zero and at its best fit value,
respectively. The fit yield is consistent with the yield ($18.2 \pm 4.8$)
obtained from the first fit shown in Fig.~\ref{fg:mpp}(b).
As a cross check, the obtained ${\mathcal B}$($J/\psi \to \pp,\LL$)
%are also shown in TABLE~\ref{br_charm}. The results 
are in good agreement with the world average
and with the latest BES result~\cite{Ablikim:2005cd}.  
We also determine the branching fraction ratios: ${\mathcal B}$($\eta_{\rm c} \to \LL$)/${\mathcal B}$($\eta_{\rm c} \to \pp$) = $0.67^{\rm+0.19}_{\rm-0.16} \pm 0.12$ 
and ${\mathcal B}$($J/\psi \to \LL$)/${\mathcal B}$($J/\psi \to \pp$) =
$0.90^{\rm+0.15}_{\rm-0.14} \pm 0.10$,
where common systematic errors in the numerator and denominator cancel.

\begin{widetext}
\begin{center}
\begin{threeparttable}
\caption{Measured branching fractions for 
$J/\psi,\eta_{\rm c} \to 
p\bar p,\Lambda\bar\Lambda$}
\label{br_charm}
\begin{tabular}{|c|c|c|c|c|}
\hline
\hline
Modes&
Yield&
Eff.(\%)&
B.F.Product$(10^{-6})$&
${\mathcal B}$($J/\psi,\eta_{\rm c} \to p\bar p,\Lambda\bar\Lambda$)$(10^{-3})$
\\
\hline
$B^+\to\eta_{\rm c}K^+,\eta_{\rm c}\to p\bar p$&
$195^{\rm +16}_{\rm -15}$&
$35.8 ^{\rm +0.3}_{\rm -0.3}$&
$1.42^{\rm +0.11}_{\rm -0.11}{+0.16\atop -0.20}$&
$1.58 \pm 0.12 ^{+0.18}_{-0.22}\pm 0.47$\tnote{a}
\\
$B^+\to\eta_{\rm c}K^+,\eta_{\rm c}\to\Lambda\bar\Lambda$&
$19.5^{\rm +5.2}_{\rm -4.5}$&
$5.3 ^{\rm +0.1}_{\rm -0.1}$&
$0.95^{\rm +0.25}_{\rm -0.22}{+0.08\atop -0.11}$ &
$0.87^{\rm+0.24}_{\rm-0.21} {+0.09\atop -0.14} \pm 0.27$\tnote{b}
\\
$B^+\to J/\psi K^+,J/\psi\to p\bar p$&
$317^{\rm +19}_{\rm -18}$&
$37.3 ^{\rm +0.4}_{\rm -0.4}$&
$2.21^{\rm +0.13}_{\rm -0.13}\pm 0.10$&
$2.21 \pm 0.13\pm 0.31\pm 0.10$\tnote{c}
\\
$B^+\to J/\psi K^+,J/\psi\to\Lambda\bar\Lambda$&
$45.9^{\rm +7.7}_{\rm -6.7}$&
$5.9^{\rm +0.3}_{\rm -0.3}$&
$2.00^{\rm +0.34}_{\rm -0.29}\pm 0.34$&
$2.00^{+0.34}_{-0.29}\pm 0.34\pm 0.08$\tnote{c}
\\
\hline
\hline
\end{tabular}
\begin{tablenotes}  
\footnotesize
\item[a] ${\mathcal B}$($B^+\to\eta_{\rm c}K^+$) = $0.9 \pm 0.27\times 10^{-3}
$\cite{PDG}. 
\item[b] We use ${\mathcal B}$($B^+\to\eta_{\rm c}K^+,\eta_{\rm c}\to\Lambda\bar\Lambda$)/${\mathcal B}$($B^+\to\eta_{\rm c}K^+,\eta_{\rm c}\to p\bar p$) =
$0.67^{\rm+0.19}_{\rm-0.16} \pm 0.12$ measured in this paper and
${\mathcal B}$($\eta_{\rm c}\to p\bar p$) = $1.3 \pm 0.4 \times 10^{-3}$
\cite{PDG}.
\item[c] ${\mathcal B}$($B^+\to J/\psi K^+$) = $1.00 \pm 0.04 \times 10^{-3}
$\cite{PDG}.
\end{tablenotes}
\end{threeparttable}
\end{center}
\end{widetext}

%We estimate the branching fraction from the ratio
%of the efficiency corrected yield of $\eta_c \to \LL$ and $\eta_c \to \pp$,
%and multiply it by $(1.3 \pm 0.4) \times 10^{-3}$ which is the ${\mathcal B}(\eta_c \to \pp)$
%PDG value~\cite{PDG}. The result is
%${\mathcal B}(\eta_c \to \LL) = ( 0.87^{+0.24}_{-0.21} (stat)^{+0.09}_{-0.14} (syst)
%\pm 0.27 ({\rm PDG}))
%\times 10^{-3}$, where the last error is associated with the 
%world average value for ${\mathcal B}(\eta_c \to \pp)$
%It is consistent with the estimation of
%. We apply the same procedure to obtain
%${\mathcal B}(J/\psi \to \LL) = ( 2.00^{+0.34}_{-0.29} (stat) \pm 0.34 (syst)
%\pm 0.08 ({\rm PDG}))
%\times 10^{-3}$ which is in good agreement with the latest BES result~\cite{Ablikim:2005cd}. 

%\begin{figure}[htb]
%\epsfig{file=fig5.eps,width=4.5in}
%\caption{$\de$ and $\mb$ distribution of
%$B^+\to\eta_{\rm c} K^+,\eta_{\rm c}\to\Lambda\bar\Lambda$
%candidates. The dark solid, solid, and dashed line represent the combined fit
%result, fitted signal, and fitted background, respectively.}
%\label{fg:etac_2dp}
%\end{figure}

Systematic uncertainties %in particle selection
are studied using high statistics control samples. For proton
identification, we use a  $\Lambda \to p \pi^-$ sample, while for
$K/\pi$ identification we use a $D^{*+} \to D^0\pi^+$,
 $D^0 \to K^-\pi^+$ sample.
The tracking efficiency is studied with
fully and partially reconstructed $D^*$ samples.
The modeling of the $\cal R$ continuum suppression requirement is studied with 
the a topologically similar control sample, $B^+ \to
J/\psi K^+, J/\psi \to \mu^+\mu^-$.
For $\Lambda$ reconstruction, we have an additional uncertainty on the
efficiency for detecting tracks away from the IP. The size of this 
uncertainty is determined from the
difference between $\Lambda$ decay-time  distributions in data and MC
simulation. 
Based on these studies,
we assign a 1\% error for each track, 2\% for each proton identification,
1\% for each kaon/pion identification, an additional 3\% for $\Lambda$
reconstruction and 3\% for the $\cal R$ selection.

The systematic uncertainty  in the fit yield is studied by varying
the parameters of the signal and background PDFs and is approximately
5\%. The MC
statistical uncertainty and modeling 
contributes a 5\% error. 
The error on the
number of $B\bar{B}$ pairs is determined to be 1\%, where
the branching fractions of $\Upsilon({\rm 4S})$ 
to neutral and charged $B\bar{B}$ pairs are assumed to be equal. 
%~\cite{Bellenote}
%Although the background in the $\mpp$ and $M_\LL$
%spectra appear negligible, 
%We force the $B$ signal yield to be 
%positive and re-fit the spectra.
The non-charmonium feed-down background below the $\eta_c$ mass region 
is estimated to be 8\% and 12\% for
the $\pp$ and $\LL$ modes, respectively.

The correlated errors are added linearly and combined quadratically with
the uncorrelated errors in the systematic error calculation. The total systematic
uncertainties are 14\% and 17\% for
the $\ppk$, and $\LLK$ modes,
respectively.

In summary, using  $386 \times 10^6$ $B\bar{B}$ events, we measure the
branching fractions of $J/\psi \to \pp$, $\eta_c \to \pp$, $J/\psi \to
\LL$ and $\eta_c \to \LL$ from $B^+ \to \ppk$ and $B^+ \to \LLK$ decays. 
%We report the first observation of $\eta_c \to \LL$ decays, with
%${\mathcal B}(\eta_c \to \LL) = ( 0.87^{+0.24}_{-0.21} \pm 0.14 \pm 0.27)
%\times 10^{-3}$.
We measure the parameter $\alpha_B$ for baryonic $J/\psi$ decays.
The parameters $\alpha_B$ are $-0.60 \pm 0.13 \pm 0.14$ and $ -0.44 \pm  0.51 \pm 
0.31$ for $J/\psi \to \pp$ and $J/\psi \to \LL$, respectively. This gives an 
$\alpha$ value for $J/\psi \to \pp$ of  $0.43 \pm 0.13 \pm 0.14$, which is 
smaller than, but
still consistent with, the current world average $0.66 \pm 0.05$.
%as shown in TABLE~\ref{alphalist}.  
%With 200 $\eta_c \to \pp$ events, we measure the decay width of
%$\eta_c$ to be . 
We also report the first observation of $\eta_c \to \LL$ decays with
${\mathcal B}(\eta_c \to \LL) = ( 0.87^{+0.24}_{-0.21} {+0.09\atop -0.14} \pm 0.27)
\times 10^{-3}$.
The observed ratio ${\mathcal B}(\eta_c \to \LL)$/${\mathcal B}(\eta_c \to \pp)$
is $0.67^{+0.19}_{-0.16} \pm 0.12$, which
is consistent with theoretical expectation~\cite{etactoBB}. 
We thank the KEKB group for excellent operation of the
accelerator, the KEK cryogenics group for efficient solenoid
operations, and the KEK computer group and
the NII for valuable computing and Super-SINET network
support.  We acknowledge support from MEXT and JSPS (Japan);
ARC and DEST (Australia); NSFC and KIP of CAS (contract No.~10575109 and IHEP-U-503, China); DST (India); the BK21 program of MOEHRD, and the
CHEP SRC and BR (grant No. R01-2005-000-10089-0) programs of
KOSEF (Korea); KBN (contract No.~2P03B 01324, Poland); MIST
(Russia); ARRS (Slovenia);  SNSF (Switzerland); NSC and MOE
(Taiwan); and DOE (USA).

\clearpage


\begin{thebibliography}{99}
\bibitem{conjugate} {Throughout this report,
inclusion of charge conjugate mode is always implied unless
otherwise stated.}
\bibitem{correction}{ There are corrections ($-2.3$ MeV, 0.5 MeV
in mean shift on $\de, \mb$ and 0.98, 1.14 in width scale on $\de, \mb$
respectively) applied to these
parameters based on the 
measured difference between data and MC simulation for $B \to D\pi$ decays.}
\end{thebibliography}
\end{document}